\documentclass[]{raa}            
\usepackage{graphicx,times}
\usepackage{natbib}
\usepackage{color}

\providecommand{\bm}[1]{\mbox{\boldmath$#1$\unboldmath}}

\def\etal{et al.}

\begin{document}

   \title{A New Multi-wavelength Solar Telescope: Optical and
   Near-infrared Solar Eruption Tracer (ONSET)
}

 \volnopage{ {\bf 2013} Vol.\ {\bf 13} No. {\bf XX}, 000--000}
   \setcounter{page}{1}

   \author{C. Fang
      \inst{1,2}
   \and P. F. Chen
      \inst{1,2}
   \and Z. Li
      \inst{1,2}
   \and M. D. Ding
      \inst{1,2}
   \and Y. Dai
      \inst{1,2}
   \and X. Y. Zhang
      \inst{3}
   \and W. J. Mao
      \inst{4}
   \and \\J. P. Zhang
      \inst{4}
   \and T. Li
      \inst{4}
   \and Y. J. Liang
      \inst{4}
   \and H. T. Lu
      \inst{4}
   }

 \institute{School of Astronomy \& Space Science, Nanjing University, Nanjing
	210093, China; {\it fangc@nju.edu.cn}\\
    \and
       Key Laboratory of Modern Astronomy \& Astrophysics,
       Nanjing University, Nanjing 210093, China
    \and
       Yunnan Astronomical Observatory, National Astronomical
       Observatories, Chinese Academy of Sciences,
       Kunming 650011, China
    \and
        Nanjing Institute of Astronomical Optics and Technology, CAS, China\\
\vs \no
   {\small Received [2013] [June] [10]; accepted [2013] [month] [day] }
}

\abstract{ A new multi-wavelength solar telescope, Optical and Near-infrared
Solar Eruption Tracer (ONSET) of Nanjing University, was constructed, being
fabricated by Nanjing Institute of Astronomical Optics \& Technology and run in
cooperation with Yunnan Astronomical Observatory. ONSET is able to observe the
Sun in three wavelength windows: He {\small I} 10830 \AA, H$\alpha$, and
white-light at 3600 \AA~ and 4250 \AA, which are selected in order to obtain
the dynamics in the corona, chromosphere, and the photosphere simultaneously.
Full-disk or partial-disk solar images with a field of 10 arcmin at three
wavelengths can be obtained nearly simultaneously. It is designed to trace
solar eruptions with high spatial and temporal resolutions. This telescope was
installed at a new solar observing site near Fuxian Lake in Yunnan Province,
southwest China. The site is located at E102N24, with an altitude of 1722 m.
The seeing is stable and of high quality. We give a brief description of the
scientific objectives and the basic structure of the ONSET. Some
preliminary results are also presented.
 }

   \authorrunning{Fang et al.}            
   \titlerunning{A new multi-wavelength solar telescope: ONSET}  
   \maketitle

%
%
\section{Introduction} \label{sect:intro}

Recently, owing to the development of solar physics and space weather 
forecast, multi-wavelength solar observations become a pressing issue 
\citep{fang11}. Fast evolution and fine structures of solar activities, such
as flares, coronal mass ejections (CMEs), and filaments, make the high 
spatio-temporal resolution observations an essential target for the
observations \citep[see, e.g.,][]{fang08}. Correspondingly, a few
new solar telescopes were contructed in the past years. For example, in
Europe, a telescope called Chromospheric Telescope (ChroTel) started to
observe the Sun in Ca {\small II} K, H$\alpha$, and He {\small I} 10830 \AA\
since April 2012 \citep{beth11}; in USA, the Synoptic Optical Long-term
Investigation of the Sun (SOLIS) was equipped with a Full-Disk Patrol (FDP)
module, which started to observe full disk images of the Sun at high cadence
in Ca {\small II} K, H$\alpha$, and He {\small I} 10830 \AA\ since June 2011
\citep{pevt11}. Therefore, there is a gap of time zone with similar
observations.

In order to fill in the gap so as to form a mini-network in monitoring the
solar chromosphere and the signatures of the corona continuously, as well as
to carry out researches on solar activities and space weather during
the solar cycle 24 in China, we decided to construct a new multi-wavelength
solar telescope called Optical \& Near-infrared Solar Eruption Tracer (ONSET),
which can conduct multi-wavelength observations in He {\small I} 10830 \AA,
H$\alpha$, and white-light at 3600 \AA~ and 4250 \AA. The operation of the
telescope is done jointly by Nanjing University and Yunnan Astronomical
Observatory. ONSET was fabricated by Nanjing Institute of Astronomical Optics
\& Technology. The project was initiated in 2005, and the telescope was
installed in 2011 at a new solar observing site at the bank of Fuxian Lake, 60
km from Kunming, Yunnan Province in southwest China. The site is located at
E102$^\circ57\arcmin11\arcsec$N24$^\circ34\arcmin47\arcsec$, with an altitude
of 1722 m. The lake is about 8 km wide and 30 km long, with the average depth
being 89.6 m. The seeing is stable, with the Fried parameter $r_0$ slightly
larger than 10 cm on average \citep{liu01}.

In this paper, we describe the scientific objectives of ONSET in \S\
\ref{sec:obj}, and the basic structure and the specifications of the telescope
are presented in \S\ \ref{sec:str}. The preliminary results are given in
\S\ \ref{sec:res}. A summary is given in \S\ \ref{sec:sum}.

\section{Scientific Objectives}\label{sec:obj}

The main objectives of ONSET are to study the following important
topics:

{\bf (1) Dynamics and fast fluctuations in flares}

Radiative transfer calculations indicate that different parts of the H$\alpha$
line may respond in different timescales to the rapid energy input
\citep{canf87}. In observations, \citet{wang00} found that the H$\alpha$--1.3
\AA\ emission in a flare shows high-frequency fluctuations on a timescale of a
few tenths of a second, which are temporally correlated with the hard X-ray
emission variations. Similar observations were done by \citet{radz11} for both
the line center and line wings. Numerical simulations by \citet{ding01} and
\citet{dingm05} verified the close relationship between the fast variations in
H$\alpha$ intensity and hard X-ray emission (the latter of which characterizes
the electron beam heating rate). Note that in some cases, the H$\alpha$ line
center is more easily affected by thermal conduction and some secondary effects
like chromospheric evaporation and Doppler shifts. Therefore, the line wings,
in particular the blue wing \citep{wang00}, is more suitable to show the
non-thermal impulsive heating.

He {\small I} 10830 \AA\ line is also a potential diagnostic tool for thermal
and nonthermal effects in solar or stellar flares \citep{ding05}. The He
{\small I} 10830 \AA\ line is a multiplet comprising of three components. This
line, although formed in upper chromospheric layers, is very sensitive to
coronal EUV irradiation. Generally speaking, there are two main ways to
populate the atomic levels responsible for this line: photoionization by EUV
irradiation followed by recombination and direct collisional excitation to the triplet levels \citep[e.g.,][]{cent08}. During solar flares, however, a third
process, collisional ionization by the electron beam followed by
recombination, may be at work \citep{ding05}. Therefore, the line formation is complicated and one should be very cautious in interpreting the line features.

Our new telescope will have a partial-disk observation mode, with a high
spatial resolution ($\sim$1\arcsec or better) and high temporal resolution
(better than 1 s), in order to study the fine structures in solar flares. This
will help reveal the physical processes regarding magnetic reconnection,
energy release, and energy transport in the solar atmosphere. In particular,
combining these data with other observations in hard X-ray, EUV, and radio,
etc., we can diagnose the thermal and non-thermal processes in flares
\citep[e.g.,][]{fang93,henoux93,henoux95} and other activities.

{\bf (2) Patrol of white-light flares}

White-light flares (WLFs) are generally thought to correspond to a type of most
energetic solar flares, with the continuum emissions from the photosphere or
lower chromosphere. They present a major challenge to the flare atmospheric
models and energy transport mechanisms \citep{neid89,ding99}. It was suggested
that all solar flares might be WLFs \citep{neid89,huds06}, which seems to be
supported by the detection of white-light emission in weak flares 
\citep{matt03,jess08, wang09} and by the spectral irradiance observations
\citep{kret11}. However, since the discovery of the first WLF in
1859 \citep{carr59}, only around 150 WLF events have been reported
conclusively so far. There are several reasons for it, e.g., (a) The
white-light enhancement in the visible continuum in a WLF is only a
few percent or less, which might below the sensitivity of some
telescopes; (b) The lifetime of the white-light emission is only
$\sim$1--2 minutes, which is too short for a certain identification.
On the other side, we have to be cautious when we use 1600 \AA\ band emissions
recorded by some spacecraft like TRACE and SDO/AIA
to search for WLFs since this band contains emission lines, e.g., C
{\small IV}. It is possible that some of the reported WLFs in 1600
\AA\ are not real WLFs.

According to whether there is a Balmer jump or not, WLFs can be
classified into type I and type II, respectively \citep{fang95}. While
type I WLFs can be explained by the standard flare model, type II
WLFs require in situ heating in the photosphere, say, low-atmosphere
magnetic reconnection \citep[e.g.,][]{li97,chen01}. Only with two
white-light wavebands inside and outside the Balmer continuum can we
distinguish which type a WLF is. For that purpose, we selected two
white-light bands, i.e., 3600 \AA~ or 4250 \AA, for the ONSET. With
the new telescope, we expect to observe more WLFs in the solar cycle
24, and to clarify some key questions regarding WLFs, such as how
common WLFs are and what heating mechanism is responsible for each of the two
types of WLFs.

{\bf (3) CME onset and filament activation}

As the largest solar eruptions which may pose hazardous impact on the
terrestrial environment or space weather, CMEs continue to attract more and
more attentions \citep{chen11}. The most important question regarding CMEs is
how the CME progenitor is triggered to erupt. Since a large fraction of CMEs
originate from filaments (or prominences when they appear above the limb), it
would be of great significance to monitor the activation of filaments. Various
observations have revealed that before a CME is formed, the associated filament
was already rising with a speed of $\sim$10 km s$^{-1}$ \citep[e.g.,][and
references therein]{cheng10}, as demonstrated by MHD numerical simulations
\citep[e.g.,][]{chen00}. The detection of such early slow rise will be
important for a better understanding of CME triggering process and for the
space weather forecast.

For the limb events, we can detect the rise motion from the imaging
observations directly. However, for the disk events, especially those near the
solar disk center, the detection of the slow rise motion relies on
spectrometers. One alternative way is to use simultaneous H$\alpha$ imaging
observations at line center and line wings (both blue and red wings). With
that, the true rising velocity of a filament on the solar disk can be measured
with a high precision \citep[e.g.,][]{mori03}.

{\bf (4) Moreton waves}

Moreton waves are fast propagating fronts observed in the chromosphere, with
an averaged velocity of $\sim$660 km s$^{-1}$ \citep{zhan11}. They typically
appear dark in the H$\alpha$ red wing and bright in the H$\alpha$ line center
and blue wing. These features can be well explained by the Uchida's model in
terms of a fast-mode mode wave in the corona sweeping the chromosphere
\citep{uchi68}. There are still many puzzles awaiting further explanations.
While Moreton waves were widely believed to be generated by the pressure pulse
in solar flares, there appeared an alternative view, i.e., they are generated
by the piston-driven shock as a flux rope erupts \citep[e.g.,][]{cliv99,
chen02}.  Another issue is the rarity of Moreton wave events. So far, it seems
that less than 40 Moreton wave events have been detected since 1960. One reason
is that most H$\alpha$ telescopes used the line center for observations,
whereas Moreton waves are better observed near H$\alpha\pm 0.45$ \AA\
\citep{chend05}.

{\bf (5) He {\small I} 10830 waves and CME nowcast}

One of the most important discoveries of the SOHO space mission is ``EIT
waves" \citep{thom98}. The intriguing phenomenon sparked world-wide debates.
A number of papers tried to explain EIT waves in terms of fast-mode waves
\citep{li12, zhen12}, there are definitely plenty of observational features
which cannot be accounted for by the fast-mode wave model, which stimulated the
birth of other non-wave models \citep[see][for a review]{chen12}. For example,
\citet{chen02} and \citet{chen05} proposed the magnetic field-line stretching
model, claiming that EIT waves are produced by the successive stretching of the
magnetic field lines straddling over the erupting flux rope. Such a model can
explain a variety of observations \citep{yang10, chen11, cheng12, dai12,
shen12}.

EIT waves were found to be the EUV counterpart of CMEs \citep{chen09}.
Therefore, their appearance can be used to nowcast CMEs, especially those
directed toward the Earth. However, EIT waves can be observed only in EUV
wavelength, which is feasible in space only. Fortunately, the coronal
brightening would result in the darkness of the near infrared line, He
{\small I} 10830 \AA\ \citep{cent08}. Therefore, even on the ground, we may
detect the near infrared counterparts of EIT waves \citep[see][for an
example]{gilb04}.

On the other hand, with the same formation mechanism of H$\alpha$ Moreton
waves, the fast-mode coronal shock wave, when sweeping the chromosphere, might
also be manifested in He {\small I} 10830 \AA. Therefore, it is possible to
detect two types of He {\small I} waves, the same as seen in EUV 
\citep{chenwu11}.

{\bf (6) Filament oscillations}

Perturbations are ubiquitous in the solar atmosphere, including the
everlasting convective motions in the photosphere, frequent brightenings in
the chromosphere, and sporadic flares in the corona. Subjected to these
perturbations, filaments would oscillate accordingly. On one hand, filament
oscillations provide an independent approach to diagnose the magnetic field
across the filaments \citep{jing03,jing06,isob06,zhan13}; On the other hand,
long-lasting filament oscillations might be one of the precursors for CME
eruptions \citep{chen08}.

{\bf (7) Other fine structures}

In the solar low atmosphere, there are a pool of small-scale
eruptive events, including microflares \citep{fang06a}, Ellerman
bombs \citep{chen01, fang06b}, and coronal bright points
\citep[CBPs,][]{zhan12}. While Ellerman bombs can be directly
observed at the wings of H$\alpha$ line, microflares and coronal
bright points are often observed in X-ray, but sometimes they show
brightenings in H$\alpha$ \citep{jian12, zhangp12}. Besides, CBPs
appear as dark points in He {\small I} 10830 \AA\ \citep{lih97}.

It is generally believed that coronal holes are the source of the fast solar
wind, which is important for space weather researches. Using He {\small I}
10830 \AA\ images, we can clearly detect the coronal holes and trace their
evolution.

\section{Basic Structure and Specifications}\label{sec:str}

\begin{figure}[h!!!]
\centerline{\includegraphics[width=9.cm,height=7.3cm]{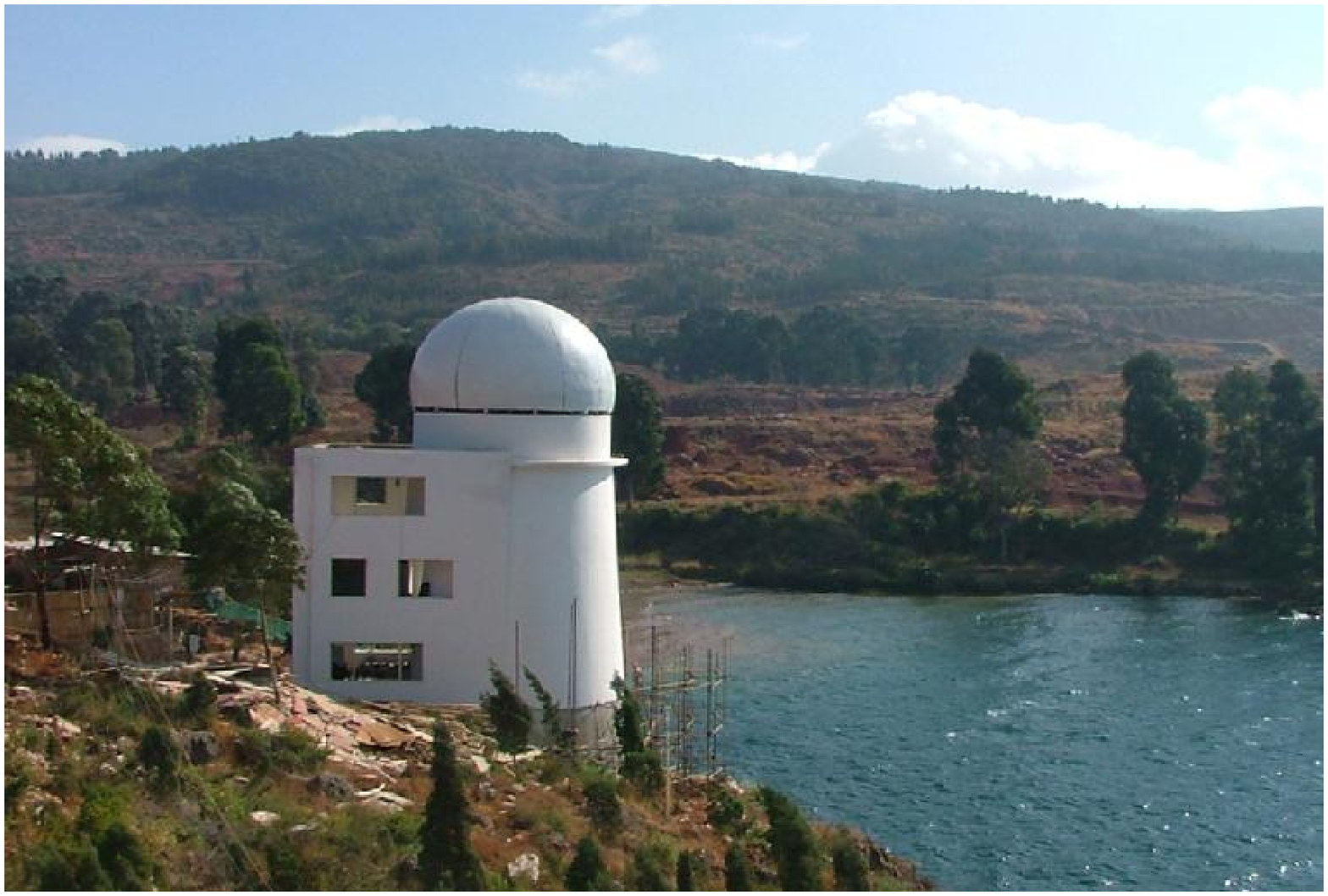} \quad
            \includegraphics[width=5.cm,height=7.3cm]{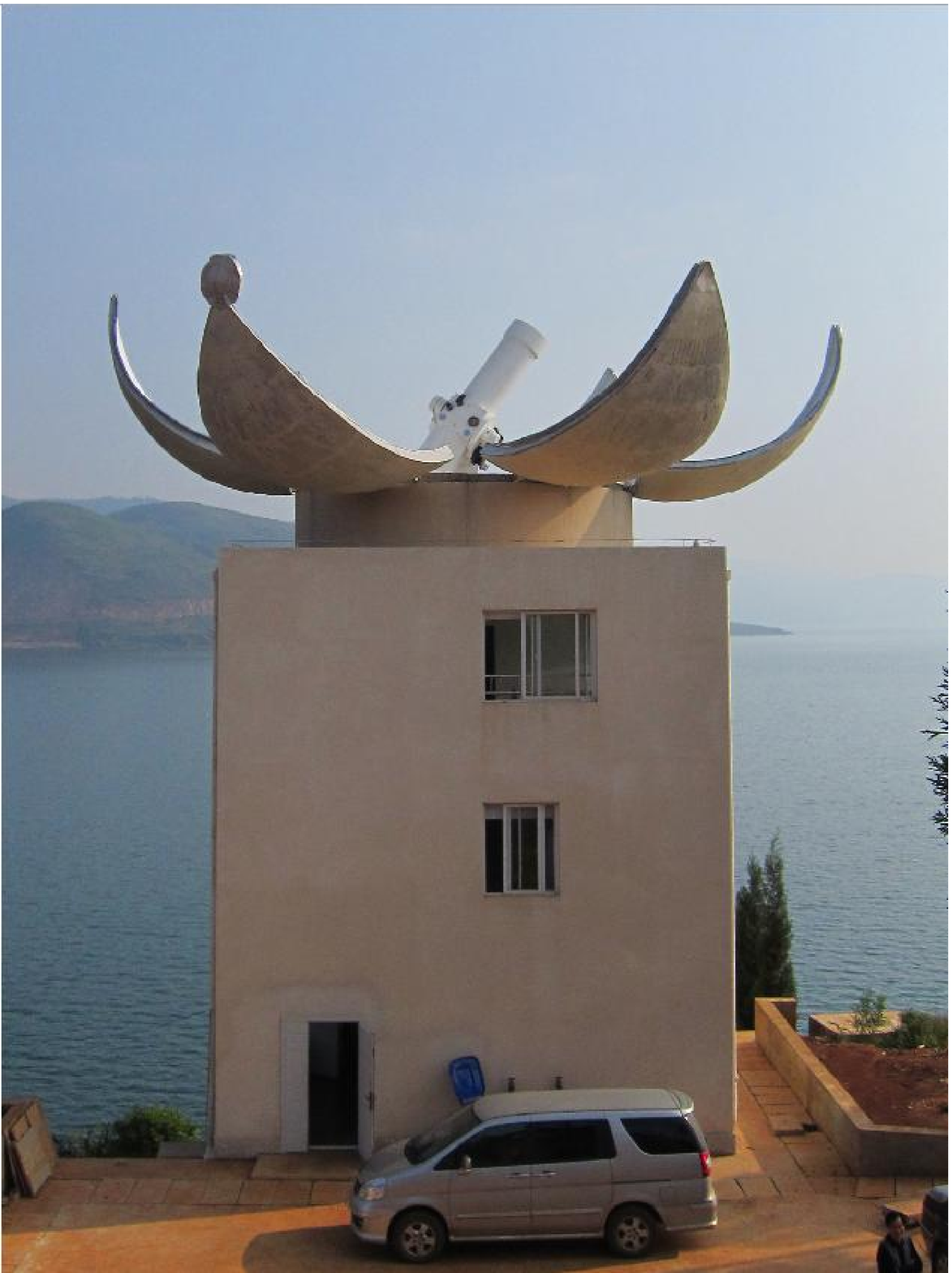}}
\caption{The ONSET building near Fuxian Lake ({\it left}) and the telescope
inside the blossom dome ({\it right}). The left picture was taken during the
construction of the building. \label{fig1}}
\end{figure}

Figure \ref{fig1} shows the ONSET building near Fuxian Lake (left panel) and
the telescope inside the blossom dome (right panel). The ONSET telescope
consists of four tubes: (1) a near-infrared vacuum tube with an aperture of
27.5 cm, observing at He {\small I} 10830$\pm$4.0 \AA\ with a FWHM of 0.5 \AA;
(2) a chromospheric (H$\alpha$) vacuum tube with an aperture of 27.5 cm,
observing at 6562.8 $\pm$ 2.5 \AA\ with a FWHM of 0.25 \AA; (3) a white-light
vacuum tube with an aperture of 20 cm, observing at the wavelength
3600 \AA~ or 4250 \AA\ with a FWHM of 15 \AA; and (4) a guiding tube
with an aperture of 14 cm, observing at the wavelength 5500 \AA.

The former two tubes are equipped with two Lyot-type filters, while
the latter two use interference filters. The first three tubes are 
vacuum ones, and they provide full-disk (extending up to 1.3$R_\odot$) 
or partial-disk images with a field of view $10\arcmin\times10\arcmin$. 
The three wavelength windows can work independently and simultaneously. Two
observation modes (full-disk and partial-disk) can be switched back and forth
according to scientific requirements. All the
observations are controlled by a computer and the observing program
is flexible. It is planned to perform routine observations to
acquire 8--10 full-disk images every 1 min in the H$\alpha$ line
center and its two wings at $\pm$0.5 \AA, He {I} 10830 \AA\ line
center and its two wings, and two white-light wavelengths. We can
also carry out partial-disk observations of solar activities with a high
spatial ($\sim$1\arcsec or better) and temporal (0.1 to 1 s)
resolutions .

The ONSET has one PIXIS 2048BR CCD for H$\alpha$, with
2048$\times$2048 pixels (the pixel size is 13.5 $\mu$m), one PCO4000
CCD for white-light observations, with 4008$\times$2672 pixels (the
pixel size is 9 $\mu$m), and one VersArray CCD for 10830 \AA~
observations, with 1340$\times$1300 pixels (the pixel size is 20
$\mu$m). The diameters of all solar images are about 24 mm at the focal plane.

\begin{figure}[h!!!]
\centerline{\includegraphics[width=12.5cm]{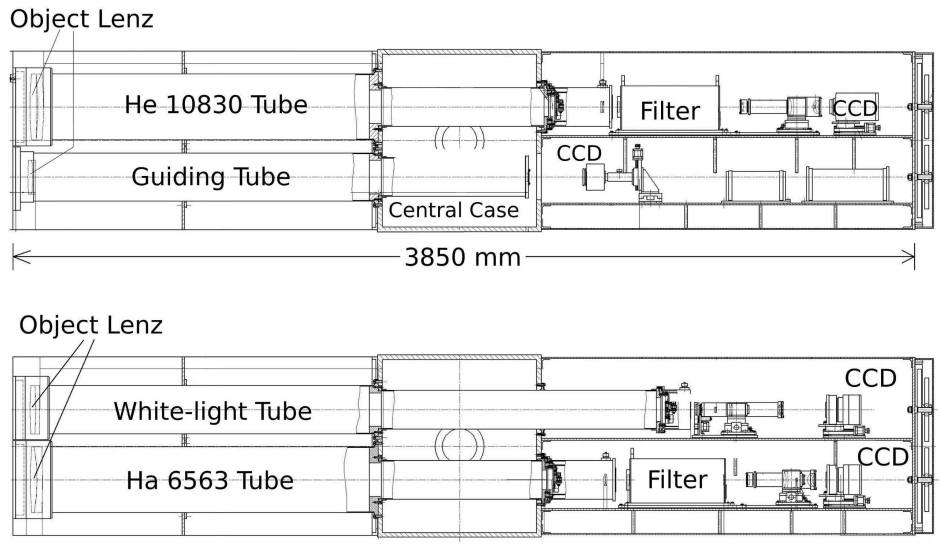}}
\caption{The inner assembly of the mechanical structure of the ONSET tubes.
The telescope consists of four tubes, including one guiding tube and three
main tubes for H$\alpha$, He {\small I} 10830 \AA, and white-light 
observations, respectively.
\label{fig2}}
\end{figure}

\begin{figure}[h!!!]
\centerline{\includegraphics[height=4.5cm]{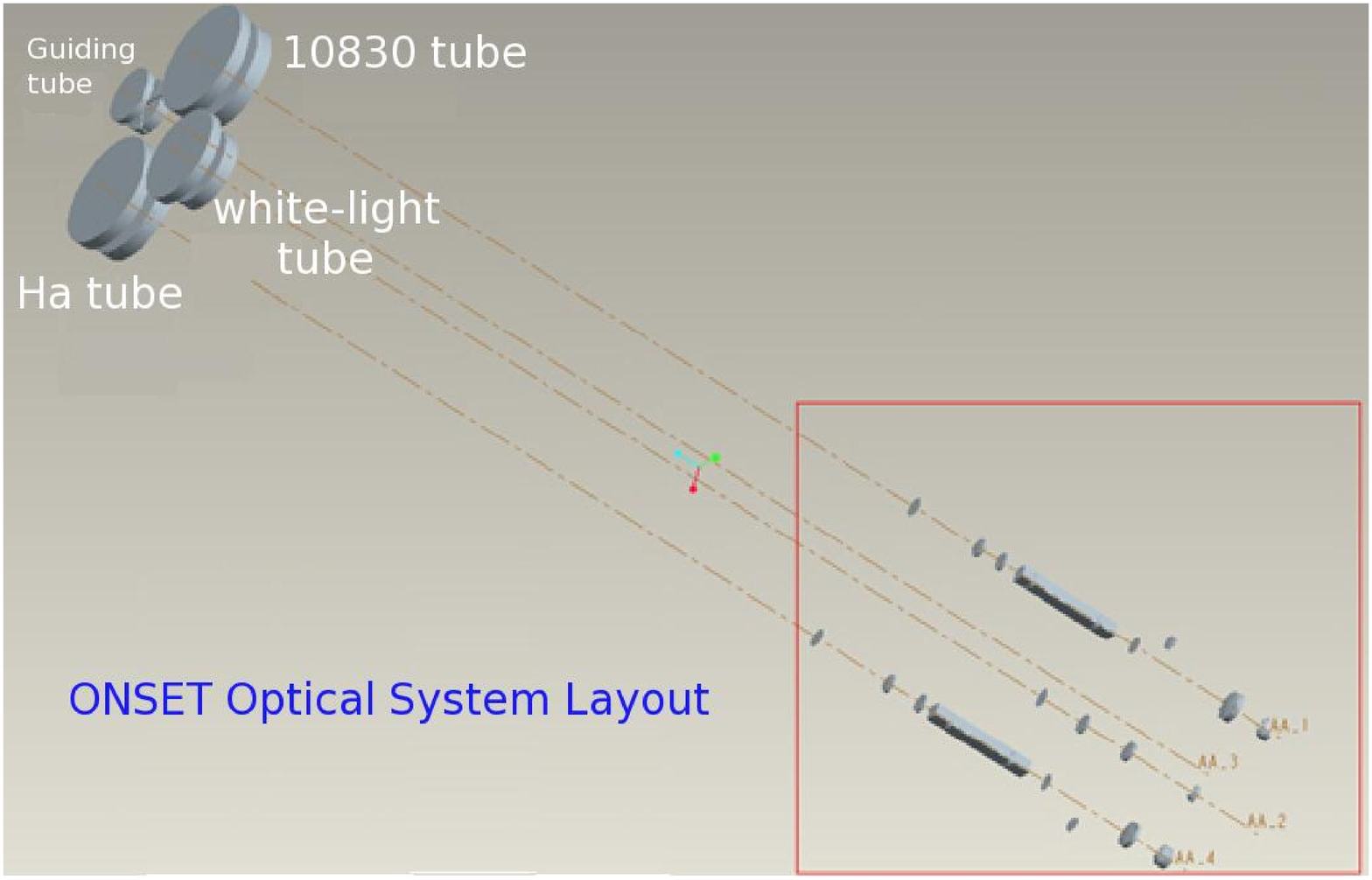} \quad
            \includegraphics[height=4.5cm]{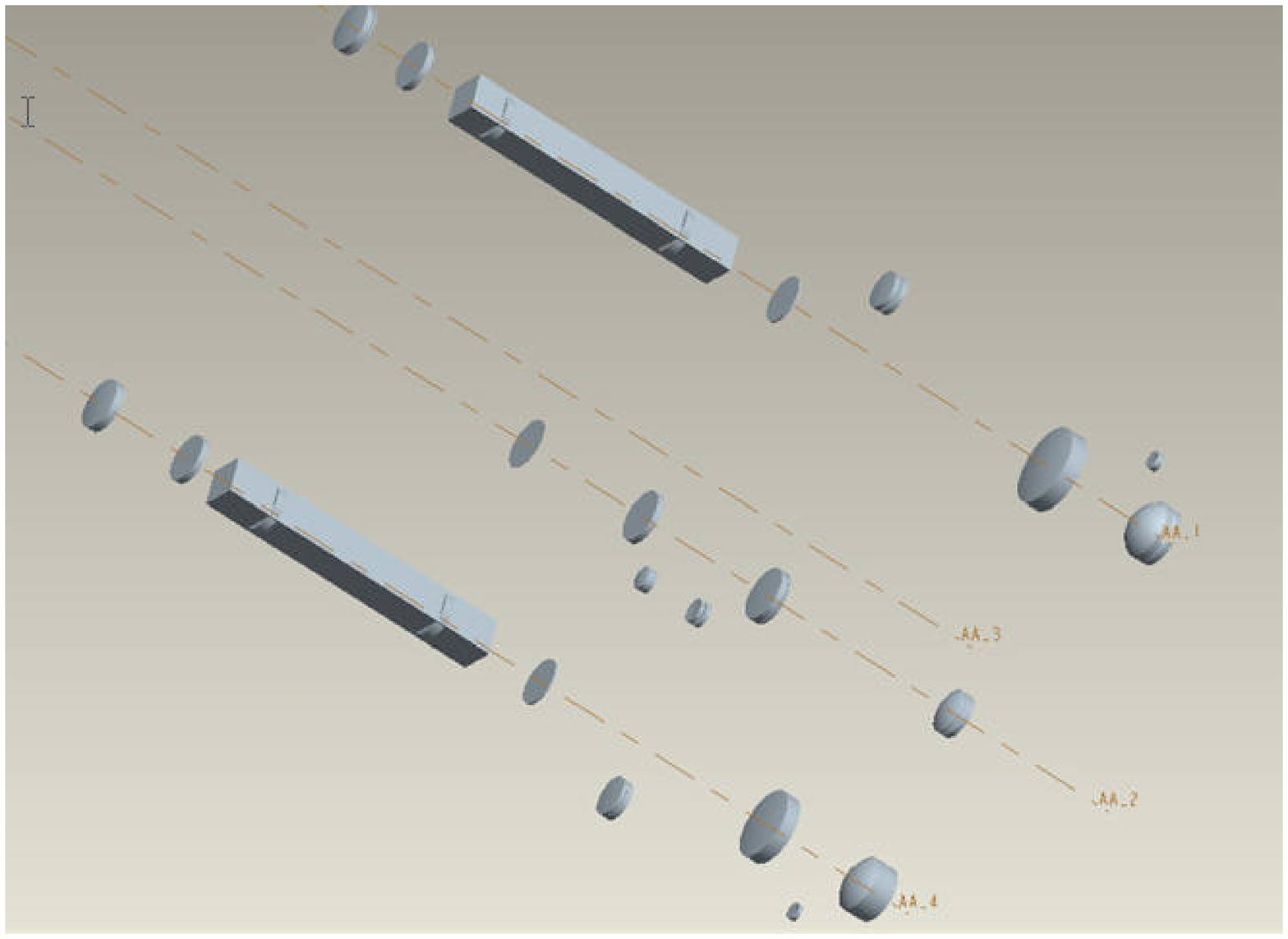}}
\caption{The optical layout of the telescope ({\it left}) and the
enlarged view of the block diagram in the left panel ({\it
right}).\label{fig3}}
\end{figure}

Figure \ref{fig2} displays the mechanical structure of the ONSET tubes, and
Figure \ref{fig3} shows the optical layout of the telescope.

A series of software programs are being developed \citep[e.g.,][]{hao13},
including data archiving, search engine, data calibrations, automatic
detections of solar activities (e.g., Moreton waves and He {\small I} 10830
waves), velocity measurements, and the remote-control system, etc.

\section{Preliminary results}\label{sec:res}

\begin{figure}[h!!!]
\centerline{\includegraphics[height=5.0cm]{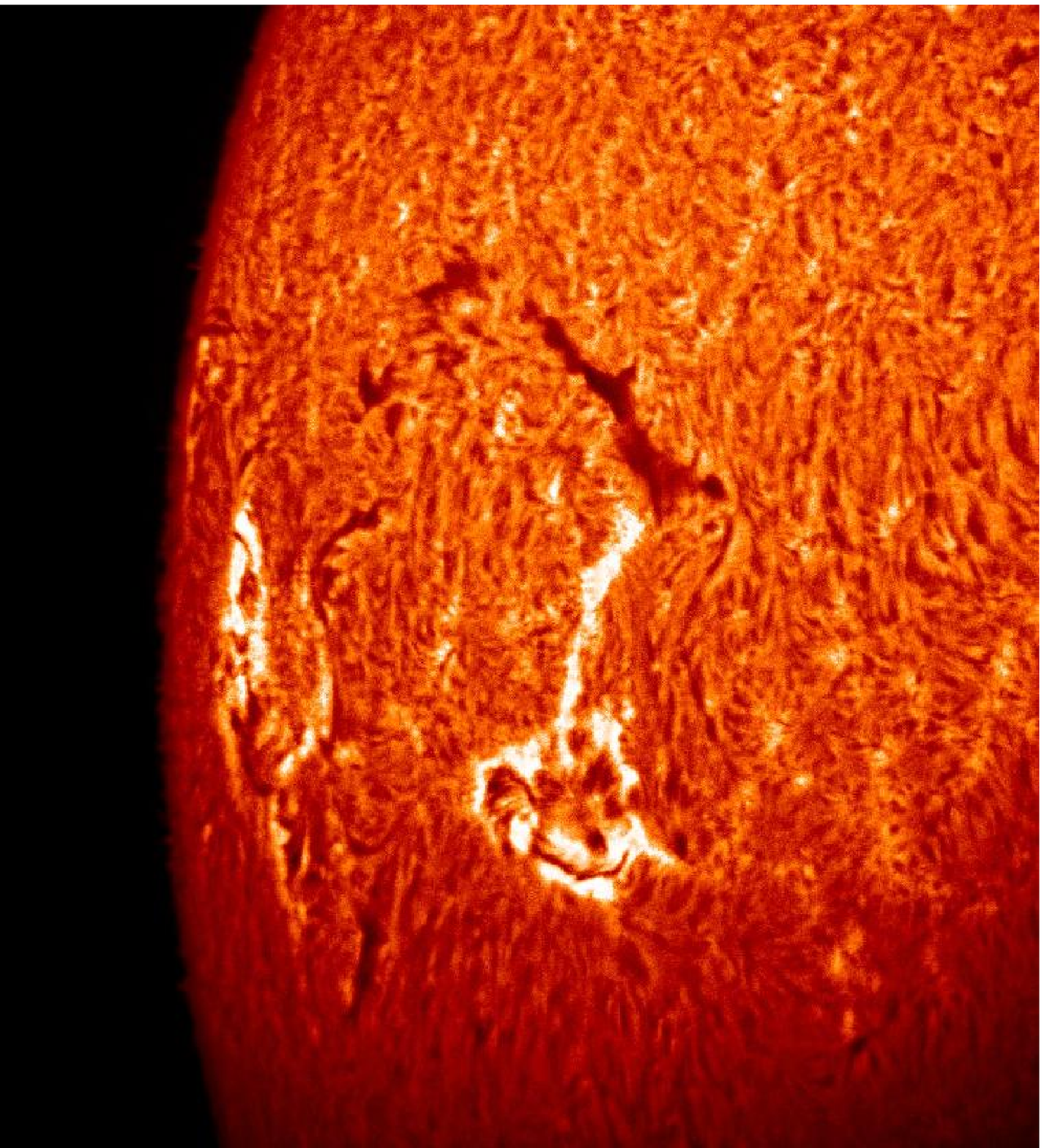} \quad
            \includegraphics[height=5.cm]{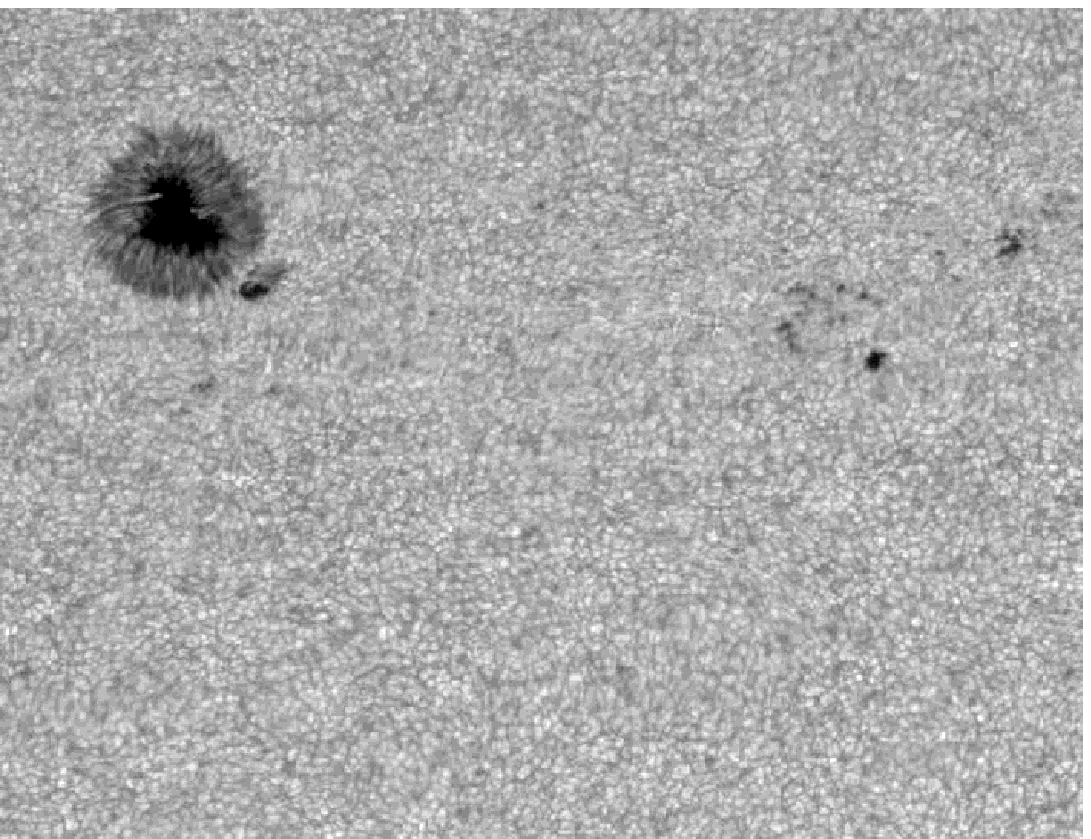}}
\caption{An H$\alpha$ partial-disk image ({\it left}) and a 4250 \AA~
partial-disk image ({\it right}) observed by ONSET. The image in the left panel
was taken at 07:05:25 UT on 21 April 2011 with a field of view of
$10\arcmin\times 10\arcmin$, whereas that in the right panel was on 14:00:52 UT
on 27 May 2013 with a field of view of $210\arcsec\times 160\arcsec$. \label{fig4}}
\end{figure}

Since 2011 April, we have made test observations of the new telescope, and got
some preliminary results. Figure \ref{fig4} shows an H$\alpha$ partial-disk
image and a white-light partial-disk image at 4250 \AA, where the fine
structures of a sunspot and the background granules are clearly visible. It
seems that the image quality is very good. In particular, a WLF was
successfully observed on 2012 March 9, which is the first flare detected by
the ONSET \citep{hao12}. Continuum emissions appear clearly at 3600 \AA~ and
4250 \AA~ wavebands. The peak enhancements at these two bands are 25\% and 
12\%, respectively. This event shows clearly the evidence that the white light
emission is caused by energetic particles bombarding the solar lower
atmosphere.

\begin{figure}[h!!!]
\centerline{\includegraphics[width=6.1cm]{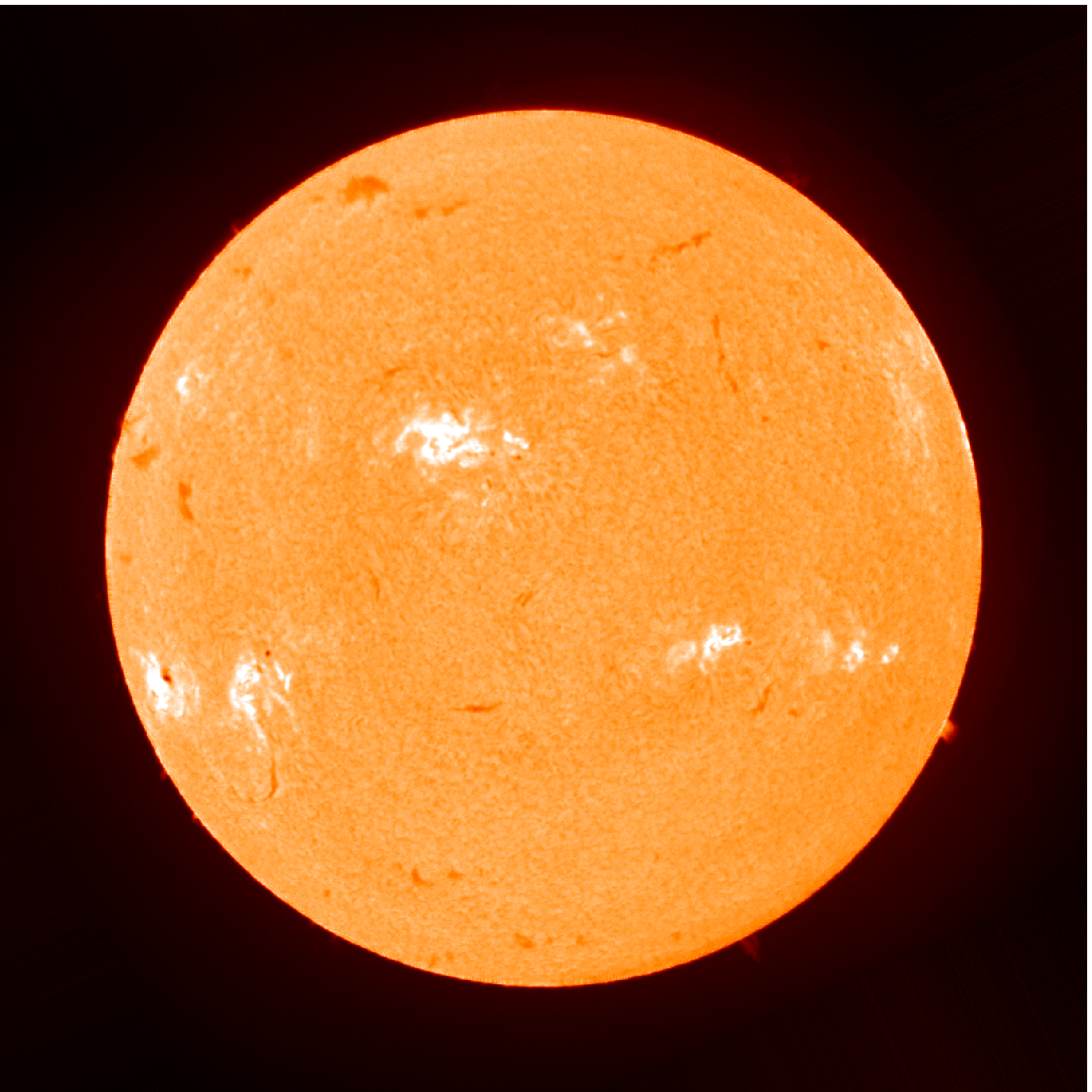} \quad
            \includegraphics[width=6.5cm]{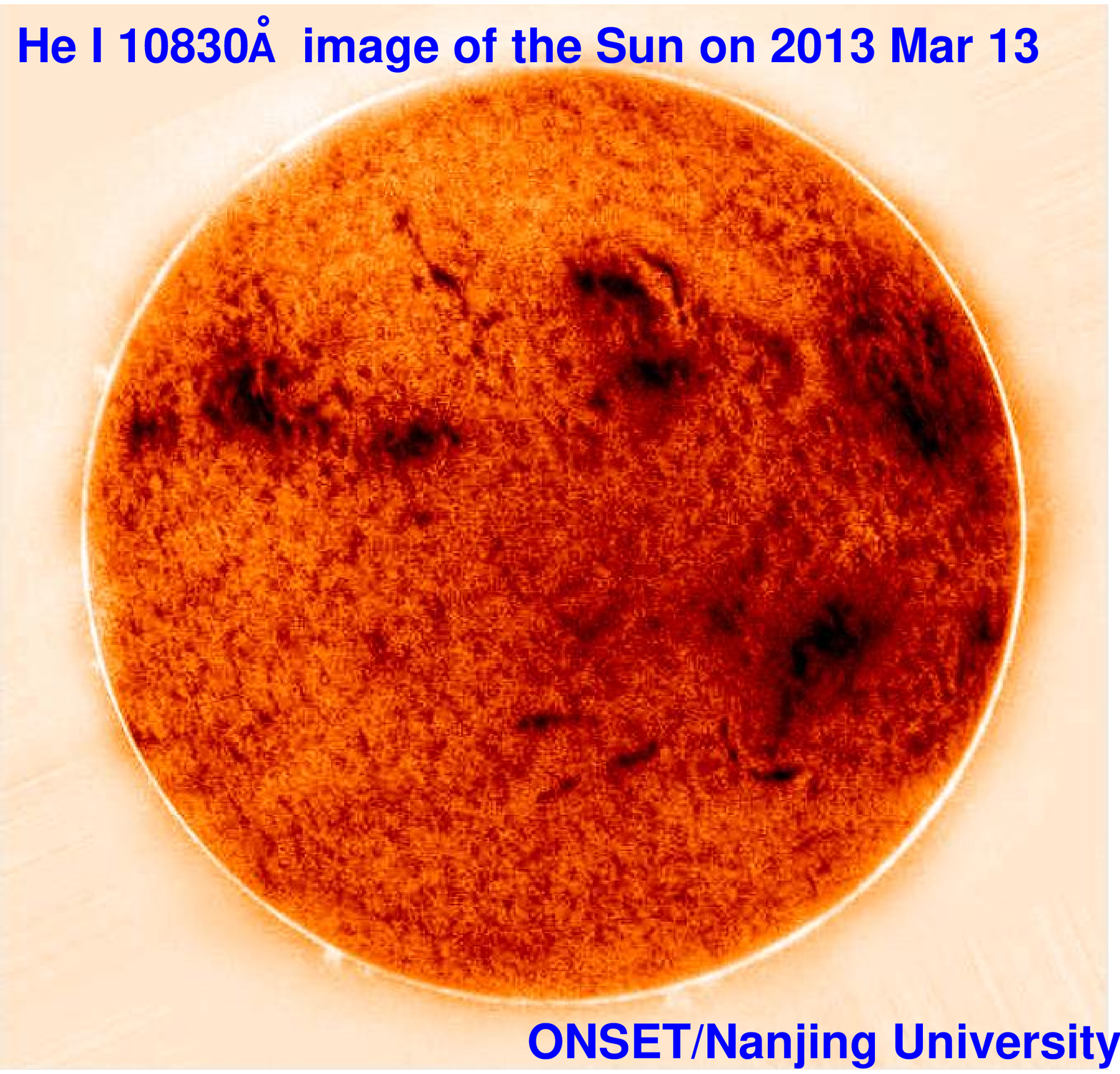}}
\caption{An H$\alpha$ ({\it left}) and a He {\small I} 10830 \AA~({\it right})
image observed by ONSET. The image in the left panel was taken at 03:09:36 UT
on 30 April 2013, whereas that in the right panel was at  03:39:21 UT on 13 March 2013. \label{fig5}}
\end{figure}

Since the beginning of 2013, ONSET has started to make routine observations.
Figure \ref{fig5} illustrates an H$\alpha$ image (left panel) and a He 
{\small I} 10830 \AA~ image (right panel), which is the first 10830 \AA\ image
observed ever in China.

ONSET is dedicated to a wide variety of solar activity researches, especially
through joint observations with other ground-based and space-born instruments.
The data from routine observations will be open and available on the internet
(\textcolor[rgb]{0.96, 0.6, 0.9}{http://sdac.nju.edu.cn}). Everyone is
welcome to use the ONSET data and to help develop software programs for the
data reduction.

\section{Summary}\label{sec:sum}

After making efforts in the past several years, we have successfully
constructed the Optical and Near-infrared Solar Eruption Tracer (ONSET) at the
new solar observing site near Fuxian Lake, which is currently the best site
for solar observations in China. The ONSET consists of four tubes, and three
of them are vacuum ones. The telescope can provide 8--10 solar images at He
{\small I} 10830 \AA\ (line center and blue/red wings), H$\alpha$ (line center
and blue/red wings), and white-light at 3600 \AA~ and 4250 \AA. The wavebands
were selected in order to observe the dynamics in the corona, chromosphere,
and the photosphere simultaneously. With the accumulation of the observational
data, it is expected to acquire the following results:

(1) Fast variations of solar flare emissions will be examined in a systematic
way in order to indirectly derive nonthermal parameters in solar flares. In
particular, the different response in H$\alpha$ and He {\small I} 10830 \AA\
to nonthermal particles can be compared;

(2) A catalog of white-light flares will be compiled, with the classification
of type I and type II events. The relationship between white-light
enhancements and soft X-ray intensity can be scrutinized;

(3) Early slow rise motion of erupting filaments can be detected with a high
cadence, which provides real-time monitoring of the onset of filament
eruptions and CMEs;

(4) Both H$\alpha$ Moreton waves and ``EIT waves" might be imprinted in
He {\small I} 10830 \AA\ images, which will help clarify the possible 
different nature of the two wavelike phenomena;

(5) Both transverse and longitudinal filament oscillations will be recorded
routinely, which paves the way to prominence seismology;

(6) A large number of small-scale eruptions will be observed, including 
Ellerman bombs and CBPs.

The preliminary results demonstrate that the quality of the solar images
is high and good enough for scientific researches. Since 2013, ONSET starts
to perform routine observations. The data will be open and available on the
internet. Everyone is welcome to use the ONSET data and to help develop
software programs.

\normalem
\begin{acknowledgements}
This work was supported by the National Natural Science Foundation of China
(NSFC) under Nos. 10878002, 10610099, 10933003, 10673004, and 11025314, as
well as the grant from the 973 project 2011CB811402 of China.
\end{acknowledgements}

\label{lastpage}

\end{document}